\documentclass[conference]{IEEEtran}
\IEEEoverridecommandlockouts

\usepackage{cite}
\usepackage{amsmath,amssymb,amsfonts}
\usepackage{algorithmic}
\usepackage{graphicx}
\usepackage{booktabs}
\usepackage{multirow}
\usepackage{textcomp}
\usepackage{xcolor}
\usepackage{balance}

\usepackage[hidelinks]{hyperref}
\def\BibTeX{{\rm B\kern-.05em{\sc i\kern-.025em b}\kern-.08em
    T\kern-.1667em\lower.7ex\hbox{E}\kern-.125emX}}

\begin{document}

\title{XFreq-GS: Cross-Frequency Wireless Radiation Field Reconstruction with 3D Gaussian Splatting}

\author{\IEEEauthorblockN{Sheng Wang\IEEEauthorrefmark{1}, Hengtao He\IEEEauthorrefmark{1}, Chaozheng Wen\IEEEauthorrefmark{2}, Jingwen Tong\IEEEauthorrefmark{3}, Xinyu Li\IEEEauthorrefmark{1}, Xiao Li\IEEEauthorrefmark{1}, Jun Zhang\IEEEauthorrefmark{2}, Shi Jin\IEEEauthorrefmark{1}}
\IEEEauthorblockA{\IEEEauthorrefmark{1}School of Information Science and Engineering, Southeast University, Nanjing 210096, China\\
}
\IEEEauthorblockA{\IEEEauthorrefmark{2}Dept. of ECE, The Hong Kong University of Science and Technology, Kowloon, Hong Kong, China
}
\IEEEauthorblockA{\IEEEauthorrefmark{3}College of Electronics and Information Engineering, Shenzhen University, Shenzhen 518060, China\\
}
\IEEEauthorblockA{E-mail: \{sheng.wang, hehengtao, xinyu\_li, li\_xiao, jinshi\}@seu.edu.cn, cwenae@connect.ust.hk,\\ eejwentong@szu.edu.cn, eejzhang@ust.hk.}}

\maketitle
\bstctlcite{BSTcontrol}

\begin{abstract}
	Channel modeling is fundamental to the analysis, design, and optimization of wireless communication systems, which, however, accurate wireless channel modeling remains challenging, especially given the increasingly complex wireless environments. As an emerging paradigm, 3D Gaussian Splatting (3DGS)-based channel modeling methods achieve accurate wireless radiation field (WRF) reconstruction and high-fidelity spatial spectrum synthesis. However, existing works only consider a single carrier frequency and fail to adapt to wide-range cross-frequency channels. To address this challenge, we propose XFreq-GS, a cross-frequency Gaussian splatting framework for WRF reconstruction. It employs 3D Gaussian primitives with shared geometry and frequency-adaptive radio frequency (RF) attributes to reconstruct cross-frequency WRF, and synthesizes power angular spectrum (PAS) maps for wireless channel modeling. Experiments show that XFreq-GS outperforms state-of-the-art 3DGS-based methods in PAS synthesis and achieves superior cross-frequency generalization. Code is available at https://github.com/KINGAZ1019/XFreq-GS.
\end{abstract}

\vspace{-6pt}
\section{Introduction}
The upcoming sixth-generation (6G) wireless networks are expected to support a wide range of emerging
applications, such as Internet of Everything, extended reality, and collaborative robotics, leading to stringent requirements on
high data rate, low latency, high reliability and connection density~\cite{alwis2021survey}. To meet these challenges
in future complex wireless environments, accurate and environment-aware wireless channel modeling is
indispensable for the analysis, design, and optimization of communication systems. By characterizing how radio signals
propagate and interact with physical environments, channel models provide essential propagation knowledge for coverage
analysis, beam management, network planning, and environment-aware
optimization~\cite{3gpp2022tr38901,almersSurveyChannelRadio2007,meinila2009winner2}.

Wireless channel modeling approaches can be broadly categorized into probabilistic, deterministic, and artificial intelligence (AI)-based
approaches. Probabilistic models, while computationally efficient, rely on empirical statistics and inherently lack
site-specific fidelity, often failing to accurately characterize detailed spatial characteristics like the angle of
arrival~\cite{almersSurveyChannelRadio2007,meinila2009winner2}. To address this, deterministic approaches, such as
ray tracing, explicitly model the electromagnetic propagation using approximated 3D
environments~\cite{seidel1994sitespecific,yun2015raytracing}. Although they provide comprehensive geometric insights, the
performance is bottlenecked by the difficulty of capturing fine-grained physical and material properties.

To address these limitations, AI has been applied to wireless channel modeling. By using advanced deep learning techniques, these
methods can reconstruct an accurate wireless radiation field (WRF) and power angular spectrum (PAS) in a data-driven
manner. Here, PAS is the receiver-side angular-domain representation of the underlying WRF. Among AI-based approaches, neural radiance field (NeRF)~\cite{mildenhall2020nerf} has shown its potential for wireless channel modeling.
NeRF-based methods learn an implicit neural representation of the wireless propagation environment, where a neural
network maps spatial positions and radio-related inputs to signal responses. Following this idea, NeRF2~\cite{zhao2023nerf2}
and NeWRF~\cite{lu2024newrf} reconstruct implicit WRF for site-specific channel modeling without explicit material modeling.
However, the inherent implicit parameterization often leads to expensive training and inference time, especially for
high-resolution spatial or angular predictions.

Recently, 3D Gaussian Splatting (3DGS)~\cite{kerbl20233dgs} stands out as a promising alternative for WRF reconstruction, inspiring several RF-oriented 3DGS
methods~\cite{wen2025wrfgs,wen2025wrfgsplus,yang2025gsrf,zhang2025rf3dgs}. For example, GSRF~\cite{yang2025gsrf} uses
complex-valued Gaussian primitives for RF data synthesis, while WRF-GS+~\cite{wen2025wrfgsplus} incorporates electromagnetic wave
propagation characteristics into Gaussian-based WRF reconstruction. These methods establish an explicit WRF reconstruction paradigm and achieve high-fidelity power angular spectrum (PAS) synthesis. However,
existing 3DGS-based methods assume single-frequency settings. They mainly work well at a single carrier frequency and cannot generalize to wide-range cross-frequency channel modeling, which severely limits their applications in future wideband communications.

To solve this problem, we propose \textbf{XFreq-GS}, a cross-frequency Gaussian splatting framework for WRF reconstruction.
We first develop a frequency-driven Gaussian representation that anchors frequency-adaptive RF attributes to
a shared scene geometry, effectively unifying multi-band wireless field modeling. Then, an RF-aware angular rendering scheme via
Adaptive Orthographic Splatting (AOS) is proposed, which directly projects the reconstructed WRF to
synthesize accurate PAS maps for channel characterization. Extensive experiments demonstrate that XFreq-GS outperforms
existing state-of-the-art 3DGS-based methods in PAS synthesis and shows strong generalization performance to different frequencies.

\section{System Model and Preliminaries}
In this section, we first introduce the frequency-dependent electromagnetic propagation effects. Based on these physical observations, we then formulate the receiver-side PAS reconstruction problem.

\begin{figure}[t]
	\centering
	\includegraphics[width=0.78\linewidth]{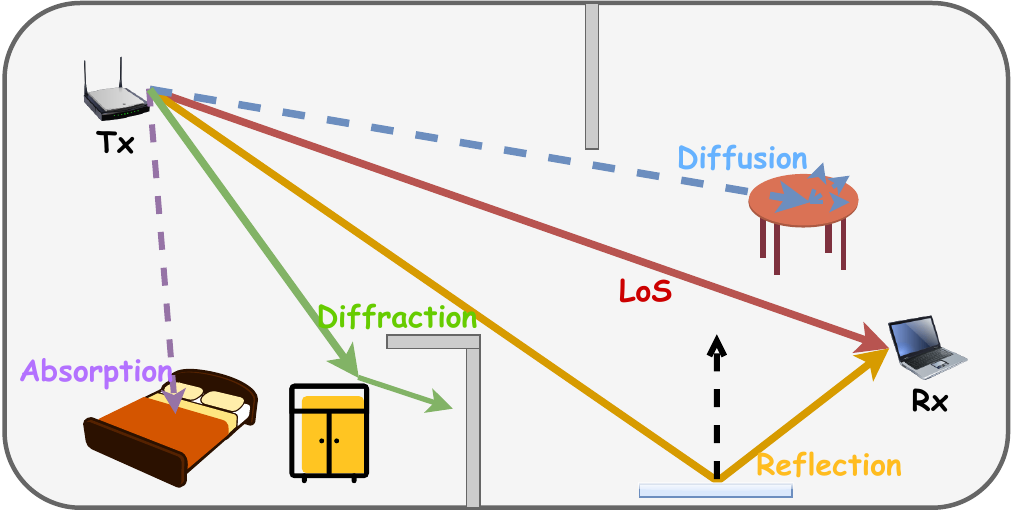}
	\caption{Illustration of wireless signal propagation in an indoor environment, where the transmitted signal reaches the receiver through multiple propagation mechanisms, including line-of-sight transmission, reflection, diffraction, diffusion, and absorption.}
	\label{fig:EM}
	\end{figure}
\subsection{Frequency-Dependent Electromagnetic Propagation}

Electromagnetic propagation critically depends on carrier frequency, as both free-space attenuation and dispersive material responses are frequency dependent~\cite{3gpp2022tr38901,yun2015raytracing,itur2023p2040}. For a fixed propagation distance $d$, the free-space path loss (FSPL) scales quadratically with the carrier frequency $f$, i.e., $\mathrm{FSPL}(d,f)=\left(4\pi d f/c\right)^{2}$, where $c$ is the speed of light. Beyond free-space propagation, indoor signals interact with surrounding objects through reflection, transmission, refraction, and scattering, as illustrated in Fig.\ref{fig:EM}. These interactions are governed by the electromagnetic properties of the material, particularly its permittivity $\varepsilon_m$ and permeability $\mu_m$.

The corresponding material impedance and free-space impedance are $\eta_m(f)=\sqrt{\mu_m(f)/\varepsilon_m(f)}$ and $\eta_0=\sqrt{\mu_0/\varepsilon_0}$, respectively. For analytical tractability, under normal incidence, the power reflection and transmission rates can be expressed as $R(f)=\left|(\eta_m(f)-\eta_0)/(\eta_m(f)+\eta_0)\right|^2$ and $T(f)=4\eta_m(f)\eta_0/(\eta_m(f)+\eta_0)^2$, with the absorption rate given by $\rho(f)=1-R(f)-T(f)$. When an EM wave passes through a material interface, the incidence angle $\theta_i$ and the refraction angle $\theta_t$ satisfy Snell's law, i.e., $\sin\theta_t/\sin\theta_i=\sqrt{\varepsilon_0\mu_0/(\varepsilon_m(f)\mu_m(f))}$.

In practice, building materials are frequency dispersive and are often expressed as $\varepsilon_m(f)$ and $\mu_m(f)$~\cite{itur2023p2040}. As a result, the macroscopic propagation mechanisms, including reflection, transmission, absorption, and angular refraction, are all coupled to the carrier frequency.

\subsection{Problem Formulation}

\begin{figure}[t]
	\centering
	\includegraphics[width=0.85\linewidth]{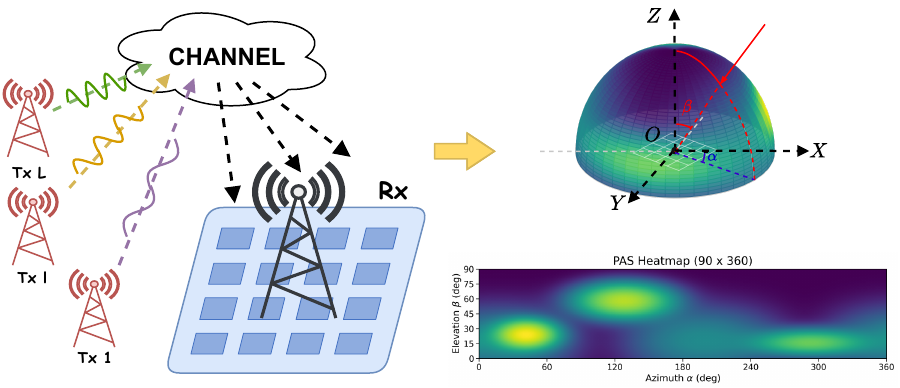}
	\caption{An illustration of wireless channel characterization through PAS modeling. Signals from multiple transmitters propagate through the wireless channel and are received by an antenna array, where the spatial distribution of received power is represented over azimuth and elevation angles.}
	\label{fig:problem_formulation}
\end{figure}

We consider a site-specific wireless propagation environment with a fixed-position receiver (RX) equipped with an antenna array and a transmitter (TX) at position $\mathbf{p}_{\mathrm{tx}}$. For carrier frequency $f$, the transmitted signal is given by $s = Ae^{j\varphi}$, where $A$ and $\varphi$ represent the signal amplitude and initial phase, respectively. After propagating through multiple effective paths induced by the frequency-dependent mechanisms discussed above, the aggregate complex signal received at the RX is given by
\vspace{-4pt}
\begin{equation}
	r
	=
	Ae^{j\varphi}
	\sum_{l=1}^{L} \Delta A_{l}
	e^{j\Delta \varphi_{l}},
	\label{eq:cf_received_signal}
\end{equation}
\vspace{-1pt}
where $L$ is the number of effective propagation paths, and $\Delta A_{l}$ and $\Delta \varphi_{l}$ denote the attenuation and phase rotation introduced by the $l$-th path, respectively.

At the RX, the multipath signal in \eqref{eq:cf_received_signal} is observed through direction-dependent phase responses across array elements. Each effective path is associated with an angle of arrival specified by its azimuth angle $\alpha$ ($0^{\circ}\leq \alpha < 360^{\circ}$) and elevation angle $\beta$ ($0^{\circ}\leq \beta < 90^{\circ}$). Following the spatial spectrum construction in multi-antenna channel modeling~\cite{wen2025wrfgs}, the RX steers narrow receiving beams toward candidate azimuth-elevation directions and measures the corresponding received power. We discretize the receiver-side hemisphere into a one-degree azimuth-elevation grid of size $90\times360$, where each grid point corresponds to one beam-steering direction. For the $(m,n)$-th angular grid point, the beamformed complex response is denoted by $y_{m,n}(\mathbf{p}_{\mathrm{tx}},f)$, and the received power at this direction is defined as
\begin{equation}
	P_{m,n}(\mathbf{p}_{\mathrm{tx}},f)
	=
	\left|
	{y}_{m,n}(\mathbf{p}_{\mathrm{tx}},f)
	\right|^{2},
	\label{eq:pas_bin}
\end{equation}
where $m$ and $n$ are the indices of the elevation and azimuth grid points, respectively. By sweeping the receiving beam over all angular grid points, we obtain the receiver-side PAS matrix
\begin{equation}
	\left[\mathbf{P}(\mathbf{p}_{\mathrm{tx}},f)\right]_{m,n}
	=
	P_{m,n}(\mathbf{p}_{\mathrm{tx}},f),
	\quad
	\mathbf{P}(\mathbf{p}_{\mathrm{tx}},f)\in \mathbb{R}_{+}^{90\times 360}.
	\label{eq:pas_matrix}
\end{equation}
The matrix $\mathbf{P}$ represents a spatial power distribution function, which directly corresponds to the WRF emitted by the transmitting source and propagated through the environment. In other words, the spatial spectrum serves as a representation of the underlying WRF, providing valuable insights into wireless channel modeling.

Fig.\ref{fig:problem_formulation} illustrates the considered fixed-RX propagation scenario, where the RX observes signals from TXs placed at different positions in a static environment. Our objective is to reconstruct receiver-side PAS for the target TX position and carrier frequency by modeling spatially distributed RF propagation responses.

\begin{figure*}[!t]
	\centering
	\includegraphics[width=\textwidth]{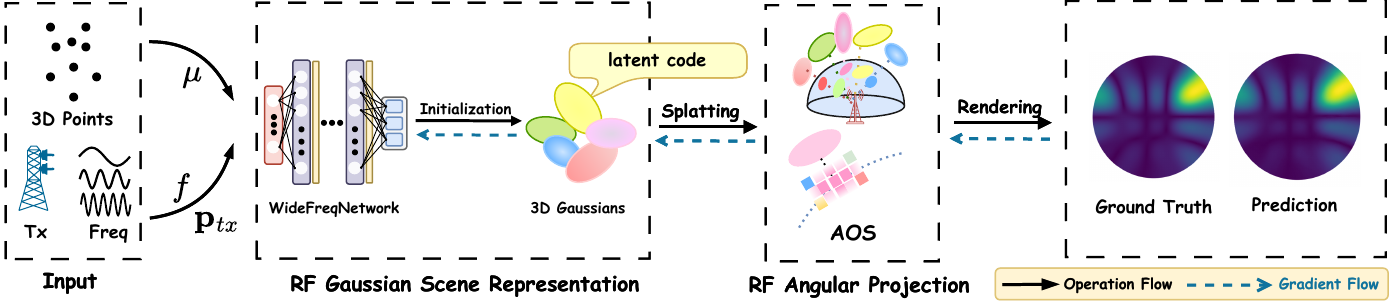}
	\caption{\textbf{Overview of the XFreq-GS pipeline for cross-frequency wireless radiation field reconstruction.} XFreq-GS
	represents the scene with RF Gaussians, where the Gaussian geometry is shared across frequencies and
	the RF attributes adapt to the TX position and carrier frequency. The RF Gaussians are then rendered onto the receiver-centered angular grid via AOS to synthesize the normalized PAS.}
	\label{fig:Overview}
\end{figure*}

\section{The Proposed XFreq-GS Framework}
In this section, we propose the XFreq-GS framework for cross-frequency WRF reconstruction using 3DGS. This framework leverages the explicit representation capability of 3DGS to characterize the spatial structure of the WRF in a given wireless environment. The key innovation lies in decoupling the frequency-invariant spatial representation from the frequency-dependent propagation characteristics.

\subsection{Overview}

As illustrated in Fig.\ref{fig:Overview}, XFreq-GS aims to synthesize the receiver-side normalized PAS for a target transmitter position $\mathbf{p}_{\mathrm{tx}}$ and carrier frequency $f$. Its key principle is that the dominant spatial support of a static environment can be shared across frequencies, whereas RF propagation attributes, such as large-scale fading, phase response, and angular energy redistribution, vary with the carrier frequency and transmitter position. XFreq-GS therefore uses the same set of 3D Gaussians as a shared geometry across frequencies, while their RF attributes are conditioned on the target TX position and carrier frequency.
This decoupling avoids learning an independent Gaussian scene for each frequency. As a result, given a set of randomly initialized 3D points, the target TX position, and the carrier frequency, XFreq-GS can construct an RF Gaussian scene and output the normalized PAS at the fixed RX antenna array. XFreq-GS consists of the following two core modules:
\begin{enumerate}
	\item \textbf{RF Gaussian Scene Representation:}   This module constructs a Gaussian-based RF scene representation from 3D points. The Gaussian geometry is shared across different frequencies, and the \textbf{WideFreqNetwork} enables each Gaussian to capture RF properties, including signal strength, attenuation, and radiation characteristics.
	\item \textbf{RF Angular Projection:} This module projects the RF Gaussians onto the receiver-centered azimuth-elevation grid and accumulates their complex contributions to generate the normalized receiver-side PAS prediction.
\end{enumerate}
In the following, we elaborate on these two modules.

\subsection{RF Gaussian Scene Representation}

Optical 3DGS represents a scene using geometry, color, and opacity~\cite{kerbl20233dgs}. In contrast, XFreq-GS assigns RF-specific attributes to each Gaussian to model signal propagation in a fixed-RX environment. Specifically, we represent the scene as a set of RF Gaussian primitives $\{\mathcal{G}_i\}_{i=1}^{N}$, where each primitive is defined by five attributes,
\begin{equation}
	\mathcal{G}_i =
	(\boldsymbol{\mu}_i,\mathbf{\Sigma}_i,\delta_i,{S}_i,\mathbf{z}_i).
	\label{eq:rf_gaussian}
\end{equation}
The first two attributes ($\boldsymbol{\mu}_i,\mathbf{\Sigma}_i$) are geometry-related. The mean $\boldsymbol{\mu}_i \in \mathbb{R}^{3}$ specifies the spatial center of the $i$-th Gaussian, while the covariance matrix $\mathbf{\Sigma}_i \in \mathbb{R}^{3 \times 3}$ determines its
geometric properties, including size, shape, and orientation. Following the standard 3DGS
parameterization~\cite{kerbl20233dgs}, we decompose the covariance matrix as
\begin{equation}
\mathbf{\Sigma}_i =
\mathbf{R}_i\mathbf{S}_i\mathbf{S}_i^{T}\mathbf{R}_i^{T},
\label{eq:covariance}
\end{equation}
where $\mathbf{R}_i$ and $\mathbf{S}_i$ denote the learnable rotation and scaling matrices, respectively. With these
parameters, the $i$-th Gaussian primitive defines an ellipsoidal spatial distribution:
\begin{equation}
G_i(\mathbf{x}) =
\exp\!\left(
-\frac{1}{2}
(\mathbf{x}-\boldsymbol{\mu}_i)^{T}
\mathbf{\Sigma}_i^{-1}
(\mathbf{x}-\boldsymbol{\mu}_i)
\right).
\label{eq:gaussian_pdf}
\end{equation}

The remaining three attributes ($\delta_i,{S}_i,\mathbf{z}_i$) are RF-related and determine how each Gaussian contributes to signal propagation. The
attenuation attribute $\delta_i$ characterizes the propagation loss introduced by the $i$-th Gaussian when RF waves
interact with the corresponding spatial region. The signal attribute $S_i$ describes the complex RF response generated
by the Gaussian, including signal strength and phase shift. The latent code $\mathbf{z}_i$ is introduced because
spherical harmonics coefficients alone cannot fully capture frequency-dependent RF effects, such as diffraction, phase interference, penetration, and frequency-selective fading. It therefore encodes local frequency-sensitive propagation variations for each Gaussian.

\begin{figure}[t]
	\centering
	\includegraphics[width=1\linewidth]{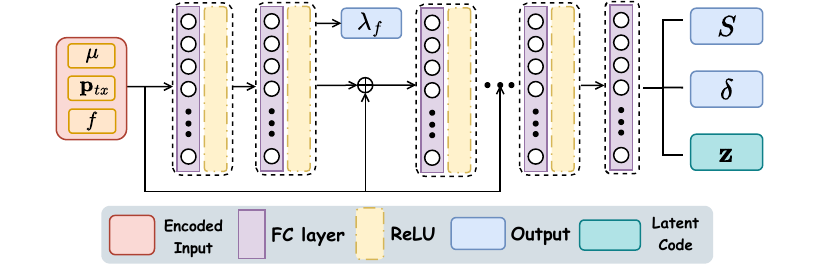}
	\caption{Architecture of WideFreqNetwork. The network encodes the TX position, Gaussian center, and carrier frequency,
	and outputs the RF-related attributes of each Gaussian, including attenuation, signal response, latent code, and angular spread factor.}
	\label{fig:network}
\end{figure}

As shown in Fig.\ref{fig:network}, WideFreqNetwork is implemented as a six-layer MLP with ReLU activations and a
hidden dimension of 256. It takes the TX position $\mathbf{p}_{\mathrm{tx}}$, Gaussian center $\boldsymbol{\mu}_i$, and carrier frequency $f$ as inputs, and outputs the attenuation, signal response, latent code, and angular spread factor of each Gaussian:
\begin{equation}
F_{\Theta}:
\left(
\mathbf{E}(\mathbf{p}_{\mathrm{tx}}),
\mathbf{E}(\boldsymbol{\mu}_i),
\mathbf{E}(f)
\right)
\Rightarrow
\left(
\delta_i,
S_i,
\mathbf{z}_i,
\lambda_{i,f}
\right),
\label{eq:widefreq_mapping}
\end{equation}
where $\Theta$ denotes the learnable network parameters, and $\mathbf{E}(\cdot)$ denotes the positional encoding
function.
Since the attenuation and signal response in RF propagation are commonly represented as complex quantities, the two
physical outputs are written as
$
\delta_i = \Delta A_i e^{j\Delta\psi_i},
S_i = A_i e^{j\psi_i},
\label{eq:complex_rf_attributes}
$
where $\Delta {A_i}$ and $\Delta\psi_i$ are the amplitude attenuation and phase shift, and $A_i$ and $\psi_i$ are the
signal amplitude and phase, respectively.

Therefore, WideFreqNetwork maps spatial Gaussian locations, TX positions, and carrier frequencies to complex RF
attributes for PAS reconstruction. This design allows XFreq-GS to handle mobile TXs and frequency-dependent propagation
effects while preserving phase information through complex-valued attenuation and signal responses. The additional
factor $\lambda_{i,f}$ controls the effective angular support used by AOS, enabling each Gaussian to adjust its
projected footprint according to the carrier frequency.

\subsection{RF Angular Projection}

After obtaining RF-related attributes, XFreq-GS renders the RF Gaussians onto the receiver-centered azimuth-elevation grid
through AOS. Unlike voxel-based RF representations that trace rays through discrete voxels, 3D Gaussians are irregularly
distributed in continuous space, making exhaustive checking between $M$ receiver-centered angular samples and $N$ RF
Gaussians costly, with $O(MN)$ operations. Inspired by 3DGS, AOS projects each RF Gaussian onto a 2D RF angular plane and localizes its influence within a finite support
region. It consists of three stages: receiver-centered angular projection, adaptive support-aware splatting, and complex
signal accumulation.

\textbf{Stage 1: Receiver-centered angular projection.}
In optical 3DGS, splatting is performed on a camera image plane. In the fixed-RX RF setting, the received PAS is
defined on a receiver-centered angular plane instead, where each sample corresponds to an azimuth-elevation receiving
direction. Let $\mathbf{r}_0$ denote the RX center and $r_{\mathrm{rx}}$ denote the radius of the receiver-centered
sampling sphere. The $i$-th Gaussian is first mapped to a receiver-centered direction and then projected onto the
2D RF angular plane:
\begin{equation}
	\hat{\mathbf{d}}_i =
	\frac{\boldsymbol{\mu}_i-\mathbf{r}_0}
	{\|\boldsymbol{\mu}_i-\mathbf{r}_0\|},
	\qquad
	\mathbf{u}_i =
	\Gamma(\mathbf{r}_0+r_{\mathrm{rx}}\hat{\mathbf{d}}_i),
	\label{eq:aos_projection}
\end{equation}
where $\hat{\mathbf{d}}_i$ is the unit direction from the RX center to Gaussian $i$, and $\Gamma(\cdot)$ maps the
corresponding point on the sampling sphere to its projected angular center $\mathbf{u}_i$.

\textbf{Stage 2: Adaptive support-aware splatting.}
The key idea of AOS is to assign each Gaussian an adaptive angular support instead of using a fixed circular footprint.
After projection, each 3D Gaussian forms an anisotropic footprint on the RF angular plane. Its shape is determined by the projected 2D covariance $\mathbf{C}_i$. The eigenvalues and eigenvectors of $\mathbf{C}
_i$ determine the footprint scale and orientation:
\begin{equation}
\mathbf{C}_i =
\mathbf{J}_i \mathbf{\Sigma}_i \mathbf{J}_i^{T},
\label{eq:aos_projected_covariance}
\end{equation}
where $\mathbf{J}_i$ is the local projection Jacobian.
To reflect frequency-dependent angular spread, we introduce a frequency-aware footprint covariance:
\begin{equation}
	\mathbf{C}_i^{f} = \lambda_{i,f} \mathbf{C}_i,
	\label{eq:aos_frequency_footprint}
\end{equation}
where $\lambda_{i,f}$ is predicted by WideFreqNetwork. A larger $\lambda_{i,f}$ produces a wider angular footprint,
while a smaller one makes the footprint more concentrated. Therefore, only angular samples inside the
adaptive footprint are affected by the Gaussian, localizing Gaussian-sample interactions while preserving
anisotropic and frequency-dependent RF propagation characteristics.

\textbf{Stage 3: Complex signal accumulation.}
For a specific angular sample $k$, AOS casts a receiver-centered angular ray and identifies the ray-wise contributing
Gaussian set whose adaptive supports cover this sample. The Gaussians in this set are ordered by their distance
from the RX. Let $N_k$ denote the size of this ordered set. The received response is obtained
by accumulating complex RF contributions along this ray:
\begin{equation}
	  R_{k} =
	  \sum_{i=1}^{N_k}
	  \mathcal{G}_i(\boldsymbol{\mu}_i,\mathbf{\Sigma}_i)
	  {S}_i(\mathbf{p}_{\mathrm{tx}},f,\mathbf{z}_i)
	  \prod_{m=1}^{i-1}
	  {\delta}_m(\mathbf{p}_{\mathrm{tx}},f,\mathbf{z}_m),
	  \label{eq:aos_complex_signal}
\end{equation}
where $R_k$ denotes the received complex response at angular sample $k$, and the index $i$ follows the receiver-centered 
distance order. The term $\mathcal{G}_i(\boldsymbol{\mu}_i,\mathbf{\Sigma}_i)$ measures the geometric
contribution of the $i$-th ray-wise Gaussian, while ${S}_i(\mathbf{p}_{\mathrm{tx}},f,\mathbf{z}_i)$ provides its
complex signal response. The product term accumulates attenuation from preceding Gaussians along the same ray. 
The receiver-side PAS is obtained from the power of the accumulated complex response and normalized to form the final PAS map. 
The predicted PAS map $\mathbf{I}_{\mathrm{pred}}$ is supervised by the ground-truth PAS map $\mathbf{I}_{\mathrm{gt}}$ through a reconstruction loss:
\begin{equation}
	  \mathcal{L}
	  =
	  (1-\lambda_{\mathrm{ssim}})\mathcal{L}_{1}
	  +
	  \lambda_{\mathrm{ssim}}\mathcal{L}_{\mathrm{ssim}},
	  \label{eq:training_objective}
\end{equation}
where $\mathcal{L}_{1}=\|\mathbf{I}_{\mathrm{pred}}-\mathbf{I}_{\mathrm{gt}}\|_{1}$ and
$\mathcal{L}_{\mathrm{ssim}}=1-\mathrm{SSIM}(\mathbf{I}_{\mathrm{pred}},\mathbf{I}_{\mathrm{gt}})$.
Here, SSIM denotes the structural similarity index measure~\cite{wang2004image}, and
$\lambda_{\mathrm{ssim}}$ balances the two loss terms. This loss optimizes the Gaussian parameters and WideFreqNetwork
parameters in an end-to-end manner.

\section{Results and Discussion}
\suppressfloats[t]

\subsection{Experimental Setup}
We evaluate XFreq-GS on simulated receiver-side PAS data under a fixed receiver configuration. The dataset~\cite{li2025wideband}
contains 14,460 samples collected from 1,446 transmitter locations at 10 specific carrier frequencies from 1~GHz to 94~GHz, where
each sample is paired with a PAS map on a $90 \times 360$ elevation-azimuth grid. We compare XFreq-GS with three
representative RF-oriented 3DGS baselines:
\begin{itemize}
	\item \textbf{WRF-GS+}~\cite{wen2025wrfgsplus}: An EM-aware 3DGS framework for WRF reconstruction with deformable
Gaussian modeling.
	\item \textbf{GSRF}~\cite{yang2025gsrf}: A complex-valued 3DGS method that models RF radiance and transmittance and
renders receiver observations through orthographic splatting.
	\item \textbf{Wideband3DGS}~\cite{li2025wideband}: A frequency-embedded RF 3DGS model that conditions Gaussian
attributes on position and carrier frequency.
\end{itemize}
For fair comparison, Wideband3DGS, WRF-GS+, and XFreq-GS use random 3D Gaussian initialization, while GSRF follows cube-sampling initialization. All methods are trained for 200k iterations using the same transmitter locations and carrier
frequencies. In addition, the quality of PAS reconstruction is evaluated by PSNR and SSIM. The cross-frequency generalization is assessed by two
experiments: leave-one-frequency-out (LOFO) evaluation and sparse-frequency interpolation.
LOFO excludes the target carrier frequency during training while
sparse-frequency interpolation tests an in-range
target frequency with limited observed training frequencies. All experiments are conducted on an NVIDIA V100 GPU.

\subsection{Performance Comparison}

\begin{figure}[!t]
	\centering
	\includegraphics[width=0.88\linewidth]{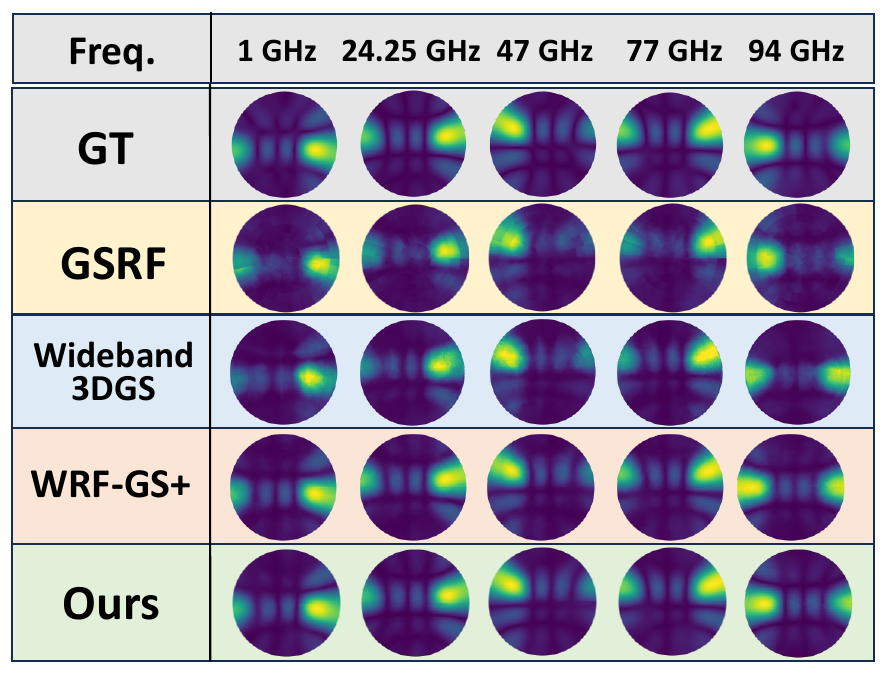}
	\caption{PAS reconstruction examples at different frequencies and TX locations.}
	\label{fig:frequency_comparison}
\end{figure}

\begin{figure}[!t]
	\centering
	\includegraphics[width=0.95\linewidth]{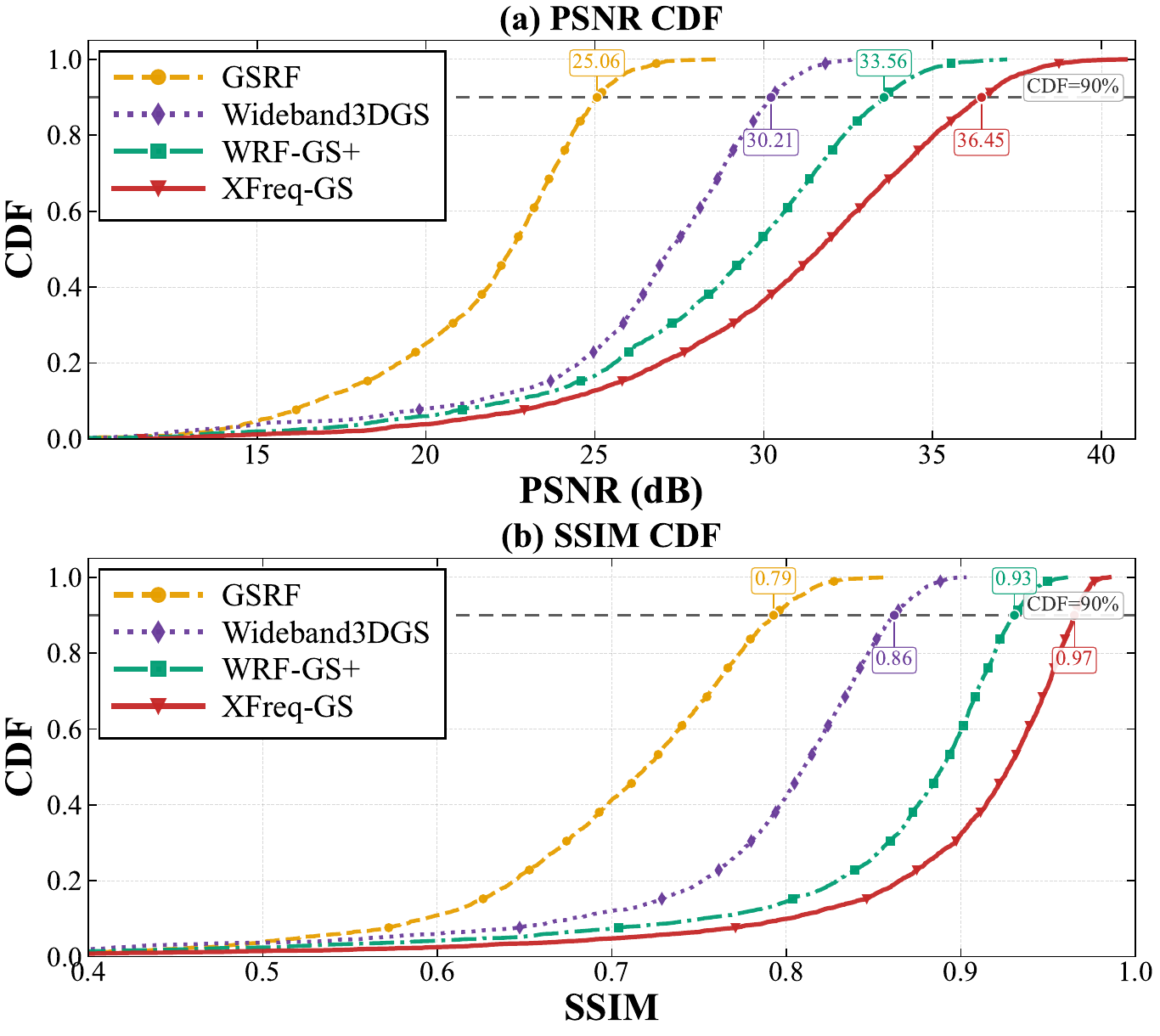}
	\caption{CDF of reconstruction performance across methods.}
	\label{fig:cdf_plot}
\end{figure}

Fig.\ref{fig:frequency_comparison} provides qualitative PAS reconstruction examples at five different carrier
frequencies and TX locations, and Fig.\ref{fig:cdf_plot} further compares the CDFs of SSIM and PSNR for different
methods. SSIM ranges from 0 to 1, with higher values indicating greater similarity between the synthesized and ground-truth samples, while higher PSNR indicates lower reconstruction error. From these results, XFreq-GS achieves the closest visual match to the ground truth and shows the best reconstruction performance. At the 90\% CDF reference level, the corresponding SSIM values are 0.97, 0.93, 0.86, and 0.79 for XFreq-GS, WRF-GS+, Wideband3DGS, and GSRF, respectively. This demonstrates that XFreq-GS can more accurately reconstruct PAS maps across different transmitter locations and carrier frequencies. The improvement comes from the proposed combination of shared scene geometry, frequency-adaptive RF attributes, and AOS, which enables XFreq-GS to model frequency-dependent attenuation and angular energy redistribution more effectively.

\subsection{Cross-Frequency Generalization}
To evaluate the unseen-frequency generalization, we conduct LOFO experiments by excluding all
samples at one target frequency from training and using them only for inference. We choose 2.4~GHz, 24.25~GHz, and
77~GHz to represent low-, mid-, and high-frequency regimes, respectively. This setting verifies whether the learned WRF
representation can transfer to entirely unseen carrier frequencies.

\begin{table}[!t]
	\centering
	\footnotesize
	\caption{Cross-Frequency LOFO Generalization Results.}
	\label{tab:crossfrequency_lofo}
	\begin{tabular}{lcccc}
	  \toprule
	  \textbf{Method} & \textbf{2.4~GHz} & \textbf{24.25~GHz} & \textbf{77~GHz} & \textbf{Avg.} \\
	  \midrule
	  GSRF & 0.7225 & 0.7227 & 0.7254 & 0.7235 \\
	  Wideband3DGS & 0.8000 & 0.7865 & 0.7966 & 0.7944 \\
	  WRF-GS+ & \underline{0.8926} & \underline{0.8709} & \underline{0.8752} & \underline{0.8796} \\
	  \textbf{XFreq-GS (Ours)} & \textbf{0.9085} & \textbf{0.8977} & \textbf{0.9232} & \textbf{0.9098} \\
	  \bottomrule
	\end{tabular}
  \end{table}

Table~\ref{tab:crossfrequency_lofo} shows the median SSIM results under the LOFO setting. XFreq-GS achieves the best median SSIM on all three unseen target bands, with values of 0.91, 0.90, and 0.92 at 2.4~GHz, 24.25~GHz, and 77~GHz, respectively. Compared with WRF-GS+, the state-of-the-art baseline, XFreq-GS improves the average SSIM by 3.4\%, with the gain increasing from 1.8\% at 2.4~GHz to 5.5\% at 77~GHz. These results indicate that XFreq-GS has better generalization performance to
unseen carrier frequencies. This advantage comes from its shared geometry and frequency-adaptive RF attributes, which
allow the learned representation to preserve common spatial structure while adapting RF responses to different carrier frequencies.

To further evaluate cross-frequency interpolation, we fix 24.25~GHz as the target frequency and vary the number of
observed training frequencies. Unlike LOFO evaluation, the target frequency lies within the observed frequency range, but
it is not included in training. This setting verifies whether XFreq-GS can reconstruct an unseen in-range frequency from sparse
frequency observations.

\begin{table}[!t]
	\centering
	\footnotesize
	\caption{Sparse-frequency interpolation at unseen $24.25\,\mathrm{GHz}$.}
	\label{tab:crossfrequency_sparse}
	\begin{tabular}{ccccc}
		\toprule
		\multirow{2}{*}{\textbf{Train Freqs. (GHz)}} & \multicolumn{2}{c}{\textbf{SSIM} $\uparrow$} & \multicolumn{2}{c}{\textbf{PSNR (dB)} $\uparrow$} \\
		\cmidrule(lr){2-3} \cmidrule(lr){4-5}
		                                         & Mean                                             & Median                                           & Mean            & Median          \\
		\midrule
		10, 37                                   & 0.7210                                           & 0.7381                                           & 22.3635         & 22.5077         \\
		10, 37, 94                   & 0.7458           & 0.7642                       & 22.9170 & 23.1264 \\
		1, 10, 37, 60, 94                        & 0.8284               & 0.8554                    & 26.5424 & 27.1833 \\
		\bottomrule
	\end{tabular}
\end{table}

Table~\ref{tab:crossfrequency_sparse} presents the sparse-frequency interpolation results at the unseen 24.25~GHz target
frequency. As the number of training frequencies increases from two to five, the performance is improved from 0.72 to 0.83
in mean SSIM and from 22.36 to 26.54~dB in mean PSNR. The median results show the same trend, increasing from 0.7381 to
0.8554 in SSIM and from 22.51 to 27.18~dB in PSNR. These results indicate that XFreq-GS benefits from denser frequency observations when reconstructing unseen in-range frequencies. They also show that additional training bands improve frequency-driven RF attribute modulation, leading to more accurate PAS reconstruction at the target frequency.

\subsection{Ablation Study}
\begin{table}[!t]
	\centering
	\footnotesize
	\caption{Ablation of XFreq-GS components.}
	\label{tab:ablation}
	\begin{tabular}{lcccc}
		\toprule
		\multirow{2}{*}{\textbf{Configuration}} & \multicolumn{2}{c}{\textbf{SSIM} $\uparrow$} & \multicolumn{2}{c}{\textbf{PSNR (dB)} $\uparrow$}                                     \\
		\cmidrule(lr){2-3} \cmidrule(lr){4-5}
		                                        & Mean                                              & Median                                       & Mean            & Median          \\
		\midrule
		Baseline                                & 0.7018                                            & 0.7204                                       & 21.7258         & 22.5307         \\
		w/o Freq.\ Modulation                   & 0.8711                                            & 0.8977                                       & 25.8631         & 26.4803         \\
		w/o AOS                                 & 0.8165                                            & 0.8360                                       & 26.2314         & 26.8358         \\
		\textbf{XFreq-GS (Full)}                    & \textbf{0.8972}                                   & \textbf{0.9278}                              & \textbf{30.7969} & \textbf{31.6460} \\
		\bottomrule
	\end{tabular}
\end{table}

Table~\ref{tab:ablation} presents the ablation study of two key designs in XFreq-GS: frequency-adaptive RF
attributes for RF Gaussian scene representation and AOS for RF angular projection. We evaluate two variants to quantify their
contributions.

Removing frequency-adaptive RF attributes reduces the mean SSIM from 0.90 to 0.87 and the mean PSNR from 30.80 to 25.86~dB. This drop indicates that fixed RF attributes fail to fully model frequency-dependent attenuation and directional response, highlighting the role of modulation in cross-frequency attribute adaptation.

Removing AOS also degrades the reconstruction quality, reducing the mean SSIM from 0.90 to 0.82 and the mean PSNR from 30.80 to 26.23~dB. The larger SSIM drop suggests that AOS is important for preserving PAS structure through receiver-centered angular projection and adaptive support.

The full XFreq-GS model achieves the best performance, with 0.90 mean SSIM and 30.80~dB mean PSNR. These results demonstrate that the two modules provide complementary benefits: frequency-adaptive RF attributes improve cross-frequency RF attribute modeling, while AOS improves angular-domain rendering fidelity. The combination enables XFreq-GS to achieve more accurate PAS reconstruction.



\section{Conclusion}
This paper proposes XFreq-GS, a cross-frequency Gaussian splatting framework for WRF reconstruction in site-specific wireless channel modeling. The core idea is to decouple frequency-shared scene geometry from frequency-adaptive RF attributes and combine this representation with AOS for RF-aware angular-domain rendering. Numerical results demonstrate that XFreq-GS consistently outperforms state-of-the-art 3DGS baselines in both PAS reconstruction and unseen-frequency generalization, while ablation results further confirm the benefits of frequency-adaptive RF attributes and AOS.
\bibliographystyle{IEEEtran}
\bibliography{IEEEabrv,references}

\begin{thebibliography}{10}
\providecommand{\url}[1]{#1}
\csname url@samestyle\endcsname
\providecommand{\newblock}{\relax}
\providecommand{\bibinfo}[2]{#2}
\providecommand{\BIBentrySTDinterwordspacing}{\spaceskip=0pt\relax}
\providecommand{\BIBentryALTinterwordstretchfactor}{4}
\providecommand{\BIBentryALTinterwordspacing}{\spaceskip=\fontdimen2\font plus
\BIBentryALTinterwordstretchfactor\fontdimen3\font minus \fontdimen4\font\relax}
\providecommand{\BIBforeignlanguage}[2]{{%
\expandafter\ifx\csname l@#1\endcsname\relax
\typeout{** WARNING: IEEEtran.bst: No hyphenation pattern has been}%
\typeout{** loaded for the language `#1'. Using the pattern for}%
\typeout{** the default language instead.}%
\else
\language=\csname l@#1\endcsname
\fi
#2}}
\providecommand{\BIBdecl}{\relax}
\BIBdecl

\bibitem{alwis2021survey}
C.~{De Alwis}, A.~Kalla, Q.-V. Pham, P.~Kumar, K.~Dev, W.-J. Hwang, and M.~Liyanage, ``Survey on {6G} frontiers: Trends, applications, requirements, technologies and future research,'' \emph{IEEE Open J. Commun. Soc.}, vol.~2, pp. 836--886, 2021.

\bibitem{3gpp2022tr38901}
{3GPP}, ``Study on channel model for frequencies from 0.5 to 100 {GHz},'' 3GPP, Technical {Report} TR 38.901 V19.3.0, Mar. 2026.

\bibitem{almersSurveyChannelRadio2007}
P.~Almers, E.~Bonek, A.~Burr, N.~Czink, M.~Debbah, V.~{Degli-Esposti}, H.~Hofstetter, P.~Ky{\"o}sti, D.~Laurenson, G.~Matz, A.~Molisch, C.~Oestges, and H.~{\"O}zcelik, ``Survey of {{Channel}} and {{Radio Propagation Models}} for {{Wireless MIMO Systems}},'' \emph{EURASIP J. Wireless Commun. Netw.}, vol. 2007, no.~1, p. 019070, Dec. 2007.

\bibitem{meinila2009winner2}
J.~Meinil{\"a}, P.~Ky{\"o}sti, T.~J{\"a}ms{\"a}, and L.~Hentil{\"a}, ``{{WINNER II}} channel models,'' in \emph{Radio Technologies and Concepts for {{IMT-Advanced}}}.\hskip 1em plus 0.5em minus 0.4em\relax Wiley, 2009, pp. 39--92.

\bibitem{seidel1994sitespecific}
S.~Y. Seidel and T.~S. Rappaport, ``Site-specific propagation prediction for wireless in-building personal communication system design,'' \emph{{IEEE} Trans. Veh. Technol.}, vol.~43, no.~4, pp. 879--891, 1994.

\bibitem{yun2015raytracing}
Z.~Yun and M.~F. Iskander, ``Ray tracing for radio propagation modeling: {{Principles}} and applications,'' \emph{{IEEE} Access}, vol.~3, pp. 1089--1100, 2015.

\bibitem{mildenhall2020nerf}
B.~Mildenhall, P.~P. Srinivasan, M.~Tancik, J.~T. Barron, R.~Ramamoorthi, and R.~Ng, ``{{NeRF}}: {{Representing}} scenes as neural radiance fields for view synthesis,'' in \emph{{Proc. Eur. Conf. Comput. Vis. (ECCV)}}, 2020, pp. 405--421.

\bibitem{zhao2023nerf2}
X.~Zhao, Z.~An, Q.~Pan, and L.~Yang, ``{{NeRF2}}: Neural radio-frequency radiance fields,'' in \emph{{Proc. ACM Annu. Int. Conf. Mobile Comput. Netw. (MobiCom)}}, 2023, pp. 393--407.

\bibitem{lu2024newrf}
H.~Lu, C.~Vattheuer, B.~Mirzasoleiman, and O.~Abari, ``{{NeWRF}}: {{A}} deep learning framework for wireless radiation field reconstruction and channel prediction,'' in \emph{{Proc. Int. Conf. Mach. Learn. (ICML)}}, 2024, pp. 33\,147--33\,159.

\bibitem{kerbl20233dgs}
B.~Kerbl, G.~Kopanas, T.~Leimk{\"u}hler, and G.~Drettakis, ``{{3D Gaussian}} splatting for real-time radiance field rendering,'' \emph{ACM Trans. Graph.}, vol.~42, no.~4, pp. 1--14, 2023.

\bibitem{wen2025wrfgs}
C.~Wen, J.~Tong, Y.~Hu, Z.~Lin, and J.~Zhang, ``{{WRF-GS}}: Wireless radiation field reconstruction with {{3D}} gaussian splatting,'' in \emph{{Proc. IEEE Conf. Comput. Commun. (INFOCOM)}}, 2025, pp. 1--10.

\bibitem{wen2025wrfgsplus}
C.~Wen, J.~Tong, Y.~Hu, Z.~Lin, and J.~Zhang, ``Neural representation for wireless radiation field reconstruction: A {{3D}} gaussian splatting approach,'' \emph{{IEEE} Trans. Wireless Commun.}, vol.~25, pp. 7490--7504, Dec. 2025.

\bibitem{yang2025gsrf}
K.~Yang, G.~Dong, S.~Ji, W.~Du, and M.~Srivastava, ``{{GSRF}}: {{Complex-valued 3D}} gaussian splatting for efficient radio-frequency data synthesis,'' in \emph{{Proc. Neural Inf. Process. Syst. (NeurIPS)}}, 2025.

\bibitem{zhang2025rf3dgs}
L.~Zhang, H.~Sun, S.~Berweger, C.~Gentile, and R.~Q. Hu, ``{{RF-3DGS}}: Wireless channel modeling with radio radiance field and {{3D}} gaussian splatting,'' \emph{{IEEE} Trans. Wireless Commun.}, vol.~25, pp. 10\,419--10\,433, 2026.

\bibitem{itur2023p2040}
\BIBentryALTinterwordspacing
{ITU-R}, ``Effects of building materials and structures on radiowave propagation in the range of 1 {MHz} to 450 {GHz},'' {International Telecommunication Union}, Tech. Rep. Recommendation {{ITU-R}} {{P}}.2040-4, Sep. 2025. [Online]. Available: \url{https://www.itu.int/rec/R-REC-P.2040-4-202509-I/en}
\BIBentrySTDinterwordspacing

\bibitem{wang2004image}
Z.~Wang, A.~C. Bovik, H.~R. Sheikh, and E.~P. Simoncelli, ``Image quality assessment: from error visibility to structural similarity,'' \emph{IEEE Transactions on Image Processing}, vol.~13, no.~4, pp. 600--612, Apr. 2004.

\bibitem{li2025wideband}
Z.~Li, L.~Yang, Y.~Bian, H.~Pan, Y.~Fu, Y.~Wang, Z.~Chen, Y.-C. Chen, and G.~Xue, ``Wideband {{RF}} radiance field modeling using frequency-embedded {{3D}} gaussian splatting,'' \emph{arXiv preprint arXiv:2505.20714}, May 2025.

\end{thebibliography}

\end{document}